\documentclass[a4paper,11pt]{article}
\usepackage{pos}

\newcommand{\be}{\begin{equation}}
\newcommand{\ee}{\end{equation}}

\newcommand{\beq}{\begin{equation}}
\newcommand{\eeq}{\end{equation}}

\newcommand\bea{\begin{eqnarray}}
\newcommand\eea{\end{eqnarray}}

\renewcommand{\thefootnote}{\alph{footnote}}

\allowdisplaybreaks
\newcommand\tb{\tan\beta}

\newcommand\ReDiag{\mathop{%
  \raise .5pt\hbox{[}%
  \widetilde{\mathrm{Re}}%
  \raise .5pt\hbox{]}}}
\newcommand\ReOffDiag{\mathop{%
  \raise .5pt\hbox{$\llbracket$}%
  \widetilde{\mathrm{Re}}%
  \raise .5pt\hbox{$\rrbracket$}}}

\newcommand\MA{M_A}

\newcommand\refta[1]{Table~\ref{#1}}

\newcommand{\sig}{\sigma}

\newcommand{\VL}{\left( \begin{array}{c}}
\newcommand{\VR}{\end{array} \right)}
\newcommand{\ML}{\left( \begin{array}{cc}}
\newcommand{\MLd}{\left( \begin{array}{ccc}}
\newcommand{\MLv}{\left( \begin{array}{cccc}}
\newcommand{\MR}{\end{array} \right)}

\definecolor{Lightblue}{cmyk}{0.9,0.1,0.1,0.3}
\definecolor{dgelborange}{cmyk}{0.,0.3,0.5, 0.}
\definecolor{Orange}{cmyk}{0.,0.5,0.5, 0.}
\definecolor{Lila}{rgb}{0.5,0.,1}

\title{Probing a Finite Unified Theory with Reduced Couplings at Future Colliders}
\author[a,b,c]{S. Heinemeyer}
\author[d]{J. Kalinowski}
\author[e]{W. Kotlarski}
\author[f]{M. Mondrag\'on}
\author*[g]{G. Patellis}
\author[g]{N. Tracas}
\author[g,h,i]{G. Zoupanos}

\affiliation[a]{Instituto de F\'{\i}sica Te\'{o}rica (UAM/CSIC), Universidad Aut\'{o}noma de Madrid,\\ Cantoblanco, 28049 Madrid, Spain}
\affiliation[b]{Campus of International Excellence UAM+CSIC, Cantoblanco, 28049 Madrid, Spain}
\affiliation[c]{Instituto de F\'{\i}sica de Cantabria (CSIC-UC), E-39005 Santander, Spain }
\affiliation[d]{University of Warsaw - Faculty of Physics,
 ul.\ Pasteura 5, 02-093 Warsaw, Poland}
\affiliation[e]{Technische Universit\"at Dresden - Institut für Kern- und Teilchenphysik (IKTP),
01069 Dresden, Germany}
\affiliation[f]{Instituto de F\'{\i}sica, Universidad Nacional Aut\'onoma de M\'exico,\\ A.P. 20-364, CDMX 01000 M\'exico}
\affiliation[g]{Physics Department, Nat. Technical University, 157 80 Zografou, Athens, Greece}
\affiliation[h]{Max-Planck Institut f\"ur Physik, F\"ohringer Ring 6, D-80805 M\"unchen, Germany}
\affiliation[i]{Theoretical Physics Department, CERN, Geneva, Switzerland}

\emailAdd{Sven.Heinemeyer@cern.ch}
\emailAdd{kalino@fuw.edu.pl}
\emailAdd{wojciech.kotlarski@tu-dresden.de}
\emailAdd{myriam@fisica.unam.mx}
\emailAdd{patellis@central.ntua.gr}
\emailAdd{ntrac@central.ntua.gr}
\emailAdd{George.Zoupanos@cern.ch}

\abstract{The search for relations among parameters that are renormalization~group invariant
to all orders in perturbation~theory constitutes the basis of the
reduction of couplings idea.~Reduction of couplings can be achieved in $N=1$ Grand Unified Theories, few of which can
become even all-loop finite. We~review the basic idea and a resulting theory in which successful~reduction of couplings has been achieved, namely the all-loop finite
$N = 1$ supersymmetric~$SU(5)$ model. We present three benchmark scenarios and~investigate~their observability at existing and future hadron colliders.~The supersymmetric~spectrum is found to be beyond the reach of the 14 TeV HL-LHC. In turn,~it is found that large parts of the predicted spectrum can be tested~at the 100~TeV FCC-hh, but the higher mass regions remain out of reach.}

\FullConference{%
  *** The European Physical Society Conference on High Energy Physics (EPS-HEP2021), ***\\
  *** 26-30 July 2021 ***\\
  *** Online conference, jointly organized by Universität Hamburg and the research center DESY ***
}

\begin{document}
\maketitle

	\section{Introduction}
The \textit{reduction of~couplings} idea
\cite{Zimmermann:1984sx}
is a promising method which relates seemingly independent parameters to a single coupling.
The~method requires the original theory to be renormalizable
and the resulting relations among parameters to be valid at all energy scales, that is 
Renormalization Group Invariant (RGI).
A next step, after the introduction of a new symmetry through a Grand Unified Theory (GUT),
in order to achieve reduction of free parameters of the Standard Model (SM), is the relation of the~gauge sector  to the~Yukawa sector. Then RGI relations (even such that can guarantee all order finiteness~of a theory) are set between the unification scale and the Planck scale.
Since supersymmetry (SUSY) seems to be a crucial ingredient  for the reduction of couplings technique to render viable phenomenology, we have to include a supersymmetry breaking sector (SSB). Significant progress~in this direction has lead to complete all-loop~finite models,~i.e. including the SSB sector. Past~applications of the above method \cite{Kapetanakis:1992vx,Kubo:1994bj} have~predicted the top quark mass one year~before its experimental measurement. Furthermore,
the all-loop finite $N=1$~$SU(5)$ model
\cite{Heinemeyer:2007tz}~has also given~a prediction for the Higgs~mass compatible with~the experimental results
and predicts a heavy SUSY mass~spectrum, consistent with the experimental~non-observation of these particles.
The full analysis of the most~successful models  can be found in recent works \cite{Heinemeyer:2020ftk,Heinemeyer:2020nzi}. 

In this article we address the question to what extent the reduction of couplings idea  can be tested at HL-LHC and FCC. To this end we propose three benchmark points for the Finite $N=1$ $SU(5)$ model and, using the SSB parameters as input in each benchmark we calculate the corresponding Higgs and SUSY spectra.~Then we compute the expected production cross sections at the 14 TeV~(HL-)LHC and the 100 TeV FCC-hh and investigate which production~channels can be observed.

%%%%%%%%%%%%%%%%%%%%%%%%%%%%%%%%%%%%%%%%%
\section{Theoretical Basis}\label{sec2}
Here we will briefly~review the core idea of the \textit{reduction of couplings} method. The target is to single out a~basic
parameter, where all other parameters can be expressed in terms of this~one
through RGI relations. Such a relation has, in general,  the~form $\Phi (g_1,\cdots,g_A) ~=~\mbox{const.}$
which should satisfy the following partial differential~equation (PDE)
\beq
\mu\,\frac{d \Phi}{d \mu} = {\vec \nabla}\Phi\cdot {\vec \beta} ~=~
\sum_{a=1}^{A}
\,\beta_{a}\,\frac{\partial \Phi}{\partial g_{a}}~=~0~,
\eeq
where $\beta_a$ is~the  $\beta$-functions of $g_a$.
The above PDE is~equivalent to the following set of ordinary differential equations (ODEs), which are called Reduction~Equations (REs)
\cite{Zimmermann:1984sx,Oehme:1984yy},
\beq
\beta_{g} \,\frac{d g_{a}}{d g} =\beta_{a}~,~a=1,\cdots,A-1~,
\label{redeq}
\eeq
where now $g$ and $\beta_g$ are~the primary coupling and its corresponding $\beta$-function.
The crucial demand is that the above~REs admit power series solutions
\beq
g_{a} = \sum_{n}\rho_{a}^{(n)}\,g^{2n+1}~,
\label{powerser}
\eeq
which preserve perturbative~renormalizability. Without this requirement, we just
trade each ``dependent'' coupling~for an integration constant. It is very important to point out that
the uniqueness of such a~solution can be already decided at the one-loop level
\cite{Zimmermann:1984sx,Oehme:1984yy}.
Concerning the reduction scheme for massive parameters,
a  number of conditions is required.~Nevertheless, progress has~been achieved, starting from \cite{Kubo:1996js}, and finally we can introduce~mass parameters and couplings carrying mass dimension
\cite{Breitenlohner:2001pp,Zimmermann:2001pq}
in the same way as~dimensionless couplings.

Consider an $N=1$  gauge theory, chiral and anomaly free,
where $G$ is the gauge group and $g$ the~associated gauge coupling. The one-loop $\beta$-function of~the gauge coupling
\cite{Parkes:1984dh,Jones:1985ay} and the one-loop anomalous dimension $\gamma^{(1)}\,^i_j$ of a chiral superfield are
\beq
\beta^{(1)}_{g}=\frac{d g}{d t} =
  \frac{g^3}{16\pi^2}\,\left[\,\sum_{i}\,T(R_{i})-3\,C_{2}(G)\,\right]~, \quad\quad\quad \gamma^{(1)}\,^i_j=\frac{1}{32\pi^2}\,\left[\,
C^{ikl}\,C_{jkl}-2\,g^2\,C_{2}(R_{i})\delta^i_j\,\right]~,
\label{betag}
\eeq
where $T(R_i)$ is the Dynkin index~of the  rep $R_i$ where the matter fields belong and $C_2(G)$~is the quadratic Casimir operator~of the adjoint rep~$G$ and $C_{ijk}$ are the Yukawa couplings.
Demanding the vanishing of all one-loop $\beta$-functions, we get
the relations
\begin{align}
\sum _i T(R_{i}) = 3 C_2(G) \,, \quad\quad\quad
 C^{ikl} C_{jkl} = 2\delta ^i_j g^2  C_2(R_i)~.
\label{1st}
\end{align}
The finiteness conditions~for an $N=1$ supersymmetric theory with $SU(N)$
associated group is found~in \cite{Rajpoot:1984zq}.
Conditions (\ref{1st}) are necessary and sufficient to
ensure two-loop finiteness
\cite{Parkes:1984dh,West:1984dg,Jones:1984cx}.
The requirement of~ one-loop finiteness in softly broken SUSY theories~demands additional constraints among the soft terms of the SSB sector~\cite{Jones:1984cu},
while, once more, these one-loop requirements~assure two-loop finiteness, too \cite{Jack:1994kd}.

The non-trivial point is that the relations among~couplings (gauge and Yukawa) which are imposed~by the~conditions (\ref{1st}) should~hold at any energy scale.
The necessary and~sufficient condition is~to
require that~such relations are solutions to the~REs 
\beq
\beta _g
\frac{d C_{ijk}}{dg} = \beta _{ijk}
\label{redeq2}
\eeq
holding at~all orders.
A theorem which points~down which are the necessary and sufficient conditions~in order for~an $N=1$ SUSY theory to be all-loop finite~can be found in 
\cite{Lucchesi:1987ef}. A comprehensive review of the above can be found in \cite{Heinemeyer:2019vbc} and \cite{Patellis:2021drd}.

%%%%%%%%%%%%%%%%%%%%%%%%%%%%%%%%%%%%%%%%%

\section{The Finite N=1 Supersymmetric SU(5) Model}\label{sec:finitesu5}

In this all-loop finite $SU(5)$ gauge theory~the reduction of couplings
is restricted to the third~generation.~An older examination of this specific Finite Unified Theory~(FUT) was shown to be in agreement with
the experimental~constraints at~the time
\cite{Heinemeyer:2007tz}~and has predicted,~almost five years before its discovery, the light~Higgs mass in the correct range. 
The particle~content of~the model has three ($\overline{\bf 5} + \bf{10}$) supermultiplets~for the three generations
of leptons~and quarks,~while the Higgs sector consists of four supermultiplets~($\overline{\bf 5} + {\bf 5}$) and one ${\bf 24}$. The~finite $SU(5)$ group~is broken to the MSSM, which of~course in no longer a~finite theory~\cite{Kapetanakis:1992vx,Kubo:1994bj,Mondragon:1993tw}.

When $SU(5)$ breaks~down to~the MSSM, a suitable rotation in the Higgs sector~\cite{Leon:1985jm,Kapetanakis:1992vx,Mondragon:1993tw,Hamidi:1984gd, Jones:1984qd},
permits~only a pair~of Higgs doublets (coupled mostly to the third family)~to remain light and acquire vev's.
Avoiding~fast proton~decay is achieved with the usual doublet-triplet splitting.~Therefore, below~the GUT scale we get the MSSM where the third generation~is given by the finiteness conditions~while the first two remain unrestricted.\\

Conditions~set by~finiteness do~not restrict the ~renormalization
properties~at low energies,~so we are left with boundary conditions
on the~gauge~and Yukawa~couplings~(plus the boundary conditions that occur in the soft sector) at~$M_{\rm GUT}$. The quark masses
$m_b (M_Z)$ and~$m_t$ are~predicted within 2$\sigma$ and
  3$\sigma$ uncertainty,~respectively, of their experimental values. 
The only~phenomenologically~viable option is to consider $\mu < 0$, as shown in~\cite{Heinemeyer:2012yj,Heinemeyer:2013fga,Heinemeyer:2018zpw,Heinemeyer:2019vbc,Heinemeyer:2020ftk,Heinemeyer:2014pbi}. The light Higgs boson mass~is predicted within 1$\sigma$ of its experimental value, while the~(point-by-point) theoretical uncertainty \cite{Bahl:2019hmm}~(calculated with {\tt FeynHiggs} \cite{Bahl:2018qog})~drops~significantly (w.r.t.~past analyses) to  $0.65-0.70$~GeV~(see \cite{Heinemeyer:2020ftk} for details).

As explained~in more~detail in~\cite{Heinemeyer:2020nzi}, the three~benchmarks chosen~(for the~purposes of collider phenomenology)~feature the LSP~above
$2100$~GeV,~$2400$~GeV~and~$2900$~GeV, respectively.
The resulting~masses~that~are relevant to our~analysis were generated by~{\tt SPheno} 4.0.4~\cite{Porod:2011nf}~and are listed in \refta{tab:futbspheno} for~each benchmark~(with the corresponding~$\tan\beta$).

%%%%%%%%%%%%%%%%%%% T A B L E %%%%%%%%%%%%%%%%%%%%%%%%%%%%%%%%%%%%%%%%%%%%%%%
\begin{center}
\begin{table}[ht]
\begin{center}
\small
\begin{tabular}{|l|r|r|r|r|r|r|r|r|r|r|r|r|}
\hline
  & $tan\beta$ & $M_{A,H}$ & $M_{H^{\pm}}$  & $M_{\tilde{g}}$ & $M_{\tilde{\chi}^0_1}$ & $M_{\tilde{\chi}^0_2}$ & $M_{\tilde{\chi}^0_3}$  & $M_{\tilde{\chi}^0_4}$ &  $M_{\tilde{\chi}_1^\pm}$ & $M_{\tilde{\chi}_2^\pm}$  \\\hline
FUTSU5-1 & 49.9 & 5.688 & 5.688  & 8.966 & 2.103 & 3.917 & 4.829 & 4.832 & 3.917 & 4.833  \\\hline
FUTSU5-2 & 50.1 & 7.039 &  7.086 & 10.380 & 2.476 & 4.592 & 5.515 & 5.518 & 4.592 & 5.519  \\\hline
FUTSU5-3 & 49.9 & 16.382 &  16.401 & 12.210 & 2.972 & 5.484 & 6.688 & 6.691 & 5.484 & 6.691  \\\hline
 & $M_{\tilde{e}_{1,2}}$ & $M_{\tilde{\nu}_{1,2}}$ & $M_{\tilde{\tau}}$ & $M_{\tilde{\nu}_{\tau}}$ & $M_{\tilde{d}_{1,2}}$ & $M_{\tilde{u}_{1,2}}$ & $M_{\tilde{b}_{1}}$ & $M_{\tilde{b}_{2}}$ & $M_{\tilde{t}_{1}}$ & $M_{\tilde{t}_{2}}$ \\\hline
FUTSU5-1 & 3.102 & 3.907 & 2.205 & 3.137 & 7.839 & 7.888 & 6.102 & 6.817 & 6.099 & 6.821 \\\hline
FUTSU5-2 & 3.623 & 4.566 & 2.517 & 3.768 & 9.059 & 9.119 & 7.113 & 7.877 & 7.032 & 7.881 \\\hline
FUTSU5-3 & 4.334 & 5.418 & 3.426 & 3.834 & 10.635 & 10.699 & 8.000 & 9.387 & 8.401 & 9.390 \\\hline
\end{tabular}
\caption{Masses~for the three~benchmarks of~the Finite $N=1$ $SU(5)$ (in~TeV)~\cite{Heinemeyer:2020nzi}.}\label{tab:futbspheno}
\end{center}
\end{table}
\end{center}
%%%%%%%%%%%%%%%%%%%%%%%%%%%%%%%%%%%%%%%%%%

At 14 TeV~HL-LHC none~of the~Finite $SU(5)$ scenarios listed above has a~SUSY~production cross section~above 0.01~fb, and thus will~most~probably~remain~unobservable~\cite{Cepeda:2019klc}.~The~discovery prospects for the~heavy Higgs-boson spectrum~is significantly better at the~FCC-hh~\cite{Hajer:2015gka}.~Theoretical analyses \cite{Craig:2016ygr,Hajer:2015gka} have~shown~that for large~$\tb$ heavy~Higgs mass~scales up to $\sim 8$ TeV~could be~accessible.~Since in this model we have $\tb \sim 50$,~the first two~benchmark points are well within the reach of~the FCC-hh~(as explained in \cite{Heinemeyer:2020nzi}).~The third~point, however, where $\MA \sim 16$~TeV,~will be far~outside~the reach~of the collider.
At 100~TeV we have in~principle production of SUSY particles in
pairs, although their~production~cross~section is~at the few fb~level.  This is a result of the~heavy~spectrum~of the model. Comparing our benchmark~predictions~with the~simplified model limits of \cite{Golling:2016gvc}, we have~found that the lighter stop might be~accessible in~FUTSU5-1~(see \cite{Heinemeyer:2020nzi}).~For~the squarks of the~first two~generations~there are better prospects.~All benchmarks~could be~tested~at~the $2\,\sig$ level,~but no discovery at~the $5\,\sig$ can~be~expected and the~same holds for the~gluino.
The expected production~cross sections for various final states and a more exhaustive discussion~can be found in \cite{Heinemeyer:2020nzi}.

%%%%%%%%%%%%%%%%%%%%%%%%%%%%%%%%%%%%%%%%%%%%%%%%%%%%%%%%%%%%%%%%%%%%%%%%%%%%%%%
%%%%%%%%%%%%%%%%%%%%%%%%%%%%%%%%%%%%%%%%%%%%%%%%%%%%%%%%%%%%%%%%%%%%%%%%%%%%%%%

\section{Conclusions}

 The reduction of couplings~scheme consists in searching for RGE
relations among parameters~of a renormalizable  theory  that hold to
all orders in perturbation~theory. In certain $N=1$ theories the
reduction of couplings idea~and the concept of finiteness are theoretically realised and~develop powerful tools able to increase the predictivity of these theories.~In the present article we turned to the question of testing~experimentally the idea of reduction of couplings. A finite model~based on the $SU(5)$ gauge theory is reviewed.

The model is found to be~in comfortable
agreement with LHC~measurements and searches, while it predicts a relatively heavy~spectrum which evades
largely the detection~in the HL-LHC.
Concerning the~accessibility of the SUSY and heavy Higgs
spectra at the~FCC-hh with $\sqrt{s} = 100$~TeV, we found that the lower parts of~the parameter
space will be~testable at the $2\,\sig$ level, with only an even smaller~part discoverable at the $5\,\sig$ level. However, the
heavier~parts of the SUSY spectrum will remain elusive even at the
FCC-hh. \\

\noindent This work has~been supported by the HARMONIA project, National Science~Center - Poland, contract UMO-2015/18/M/ST2/00518 and by the Basic~Research Programme, PEVE2020 of the National Technical University~of Athens, Greece. The work of M.M. is supported by a DGAPA-UNAM~grant PAPIIT IN109321.

\bibliographystyle{h-physrev5}
\bibliography{biblio}

\begin{thebibliography}{10}

\bibitem{Zimmermann:1984sx}
W.~Zimmermann,
\newblock Commun. Math. Phys. {\bf 97}, 211 (1985).
%%CITATION = CMPHA,97,211;%%

\bibitem{Kapetanakis:1992vx}
D.~Kapetanakis, M.~Mondragon, and G.~Zoupanos,
\newblock Z. Phys. {\bf C60}, 181 (1993), arXiv:hep-ph/9210218.
%%CITATION = HEP-PH/9210218;%%

\bibitem{Kubo:1994bj}
J.~Kubo, M.~Mondragon, and G.~Zoupanos,
\newblock Nucl. Phys. {\bf B424}, 291 (1994).
%%CITATION = NUPHA,B424,291;%%

\bibitem{Heinemeyer:2007tz}
S.~Heinemeyer, M.~Mondragon, and G.~Zoupanos,
\newblock JHEP {\bf 07}, 135 (2008), arXiv:0712.3630.
%%CITATION = 0712.3630;%%

\bibitem{Heinemeyer:2020ftk}
S.~Heinemeyer, M.~Mondragón, G.~Patellis, N.~Tracas, and G.~Zoupanos,
\newblock Fortsch. Phys. {\bf 68}, 2000028 (2020), arXiv:2002.10983.
%%CITATION = ARXIV:2002.10983;%%

\bibitem{Heinemeyer:2020nzi}
S.~Heinemeyer {\em et~al.},
\newblock Eur. Phys. J. {\bf C81}, 185 (2021), arXiv:2011.07900.
%%CITATION = ARXIV:2011.07900;%%

\bibitem{Oehme:1984yy}
R.~Oehme and W.~Zimmermann,
\newblock Commun. Math. Phys. {\bf 97}, 569 (1985).
%%CITATION = CMPHA,97,569;%%

\bibitem{Kubo:1996js}
J.~Kubo, M.~Mondragon, and G.~Zoupanos,
\newblock Phys. Lett. {\bf B389}, 523 (1996), arXiv:hep-ph/9609218.
%%CITATION = HEP-PH/9609218;%%

\bibitem{Breitenlohner:2001pp}
P.~Breitenlohner and D.~Maison,
\newblock Commun. Math. Phys. {\bf 219}, 179 (2001).
%%CITATION = CMPHA,219,179;%%

\bibitem{Zimmermann:2001pq}
W.~Zimmermann,
\newblock Commun.Math.Phys. {\bf 219}, 221 (2001).
%%CITATION = CMPHA,219,221;%%

\bibitem{Parkes:1984dh}
A.~Parkes and P.~C. West,
\newblock Phys. Lett. {\bf B138}, 99 (1984).
%%CITATION = PHLTA,B138,99;%%

\bibitem{Jones:1985ay}
D.~R.~T. Jones and A.~J. Parkes,
\newblock Phys. Lett. {\bf B160}, 267 (1985).
%%CITATION = PHLTA,B160,267;%%

\bibitem{Rajpoot:1984zq}
S.~Rajpoot and J.~G. Taylor,
\newblock Phys. Lett. {\bf B147}, 91 (1984).
%%CITATION = PHLTA,B147,91;%%

\bibitem{West:1984dg}
P.~C. West,
\newblock Phys. Lett. {\bf B137}, 371 (1984).
%%CITATION = PHLTA,B137,371;%%

\bibitem{Jones:1984cx}
D.~R.~T. Jones and L.~Mezincescu,
\newblock Phys. Lett. {\bf B138}, 293 (1984).
%%CITATION = PHLTA,B138,293;%%

\bibitem{Jones:1984cu}
D.~R.~T. Jones, L.~Mezincescu, and Y.~P. Yao,
\newblock Phys. Lett. {\bf B148}, 317 (1984).
%%CITATION = PHLTA,B148,317;%%

\bibitem{Jack:1994kd}
I.~Jack and D.~R.~T. Jones,
\newblock Phys. Lett. {\bf B333}, 372 (1994), hep-ph/9405233.
%%CITATION = HEP-PH/9405233;%%

\bibitem{Lucchesi:1987ef}
C.~Lucchesi, O.~Piguet, and K.~Sibold,
\newblock Phys. Lett. {\bf B201}, 241 (1988).
%%CITATION = PHLTA,B201,241;%%

\bibitem{Heinemeyer:2019vbc}
S.~Heinemeyer, M.~Mondragón, N.~Tracas, and G.~Zoupanos,
\newblock Phys. Rept. {\bf 814}, 1 (2019), arXiv:1904.00410.
%%CITATION = ARXIV:1904.00410;%%

\bibitem{Patellis:2021drd}
G.~Patellis,
\newblock PhD thesis, Natl. Tech. U., Athens, 2020, arXiv:2102.01476.
%%CITATION = ARXIV:2102.01476;%%

\bibitem{Mondragon:1993tw}
M.~Mondragon and G.~Zoupanos,
\newblock Nucl. Phys. Proc. Suppl. {\bf 37C}, 98 (1995).
%%CITATION = NUPHZ,37C,98;%%

\bibitem{Leon:1985jm}
J.~Leon, J.~Perez-Mercader, M.~Quiros, and J.~Ramirez-Mittelbrunn,
\newblock Phys. Lett. {\bf B156}, 66 (1985).
%%CITATION = PHLTA,B156,66;%%

\bibitem{Hamidi:1984gd}
S.~Hamidi and J.~H. Schwarz,
\newblock Phys. Lett. {\bf B147}, 301 (1984).
%%CITATION = PHLTA,B147,301;%%

\bibitem{Jones:1984qd}
D.~R.~T. Jones and S.~Raby,
\newblock Phys. Lett. {\bf B143}, 137 (1984).
%%CITATION = PHLTA,B143,137;%%

\bibitem{Heinemeyer:2012yj}
S.~Heinemeyer, M.~Mondragon, and G.~Zoupanos,
\newblock Phys.Lett. {\bf B718}, 1430 (2013), arXiv:1211.3765.
%%CITATION = ARXIV:1211.3765;%%

\bibitem{Heinemeyer:2013fga}
S.~Heinemeyer, M.~Mondragon, and G.~Zoupanos,
\newblock Phys.Part.Nucl. {\bf 44}, 299 (2013).
%%CITATION = PPNUE,44,299;%%

\bibitem{Heinemeyer:2018zpw}
S.~Heinemeyer, M.~Mondragón, G.~Patellis, N.~Tracas, and G.~Zoupanos,
\newblock Symmetry {\bf 10}, 62 (2018), arXiv:1802.04666.
%%CITATION = ARXIV:1802.04666;%%

\bibitem{Heinemeyer:2014pbi}
S.~Heinemeyer, M.~Mondrag\'on, N.~D. Tracas, and G.~Zoupanos,
\newblock Ann. U. Craiova Phys. {\bf 24}, 56 (2014).

\bibitem{Bahl:2019hmm}
H.~Bahl, S.~Heinemeyer, W.~Hollik, and G.~Weiglein,
\newblock Eur. Phys. J. {\bf C80}, 497 (2020), arXiv:1912.04199.
%%CITATION = ARXIV:1912.04199;%%

\bibitem{Bahl:2018qog}
H.~Bahl {\em et~al.},
\newblock Comput. Phys. Commun. {\bf 249}, 107099 (2020), arXiv:1811.09073.

\bibitem{Porod:2011nf}
W.~Porod and F.~Staub,
\newblock Comput. Phys. Commun. {\bf 183}, 2458 (2012), arXiv:1104.1573.
%%CITATION = ARXIV:1104.1573;%%

\bibitem{Cepeda:2019klc}
M.~Cepeda {\em et~al.},
\newblock CERN Yellow Rep. Monogr. {\bf 7}, 221 (2019), arXiv:1902.00134.

\bibitem{Hajer:2015gka}
J.~Hajer, Y.-Y. Li, T.~Liu, and J.~F.~H. Shiu,
\newblock JHEP {\bf 11}, 124 (2015), arXiv:1504.07617.

\bibitem{Craig:2016ygr}
N.~Craig, J.~Hajer, Y.-Y. Li, T.~Liu, and H.~Zhang,
\newblock JHEP {\bf 01}, 018 (2017), arXiv:1605.08744.

\bibitem{Golling:2016gvc}
T.~Golling {\em et~al.},
\newblock (2016), arXiv:1606.00947.

\end{thebibliography}

\end{document}